\documentclass[11pt]{article}
\usepackage{graphicx,color} 

\usepackage{fancyhdr}
\usepackage{isomath}
\usepackage{mathtools} 
\usepackage{amsbsy}
\usepackage{amssymb}
\usepackage{amscd,amsfonts}
\usepackage{bigints}
\usepackage{graphicx}
\usepackage{verbatim}
\usepackage{euscript}
\usepackage{alltt}
\usepackage{stmaryrd}
\usepackage{relsize}
\usepackage{enumerate}
\usepackage{url}
\usepackage[stable]{footmisc}
\usepackage{breakurl}
\usepackage{hyperref}
\usepackage{comment}

\usepackage[maxbibnames=9,sorting=none]{biblatex}
\usepackage[font=small,labelfont=bf]{caption}
\usepackage[font=small,labelfont=bf]{subcaption}
\usepackage{float}
\usepackage{mwe}
\usepackage{algorithm}
\usepackage{algpseudocode}
\usepackage{xcolor}
\usepackage[super]{nth}
\usepackage{soul}

\DeclareGraphicsExtensions{.eps,.pdf}
\usepackage{amsmath}
\usepackage{amsbsy}
\usepackage{amssymb}
\usepackage{amscd}
\usepackage{amsfonts}
\usepackage{isomath}

\newcommand{\R}{\mathbb R}

\newcommand{\C}{\mathbb C}

\newcommand{\bfg}{{\mathbold g}}

\newcommand{\bfn}{{\mathbold n}}

\newcommand{\bfu}{{\mathbold u}}
\newcommand{\bfv}{{\mathbold v}}

\newcommand{\bfE}{{\mathbold E}}

\newcommand{\bfT}{{\mathbold T}}
\newcommand{\bfU}{{\mathbold U}}
\newcommand{\bfV}{{\mathbold V}}

\newcommand{\bfX}{{\mathbold X}}

\newcommand{\eps}{{\varepsilon}}

\newcommand{\beq}{\begin{equation}}
\newcommand{\eeq}{\end{equation}}
\newcommand{\beqs}{\begin{eqnarray}}
\newcommand{\eeqs}{\end{eqnarray}}
\newcommand{\beql}{\begin{equation} \label}
\newcommand{\half}{\frac{1}{2}}

\newcommand{\bfalpha}{\mathbold{\alpha}}

\newcommand{\bfzero}{\mathbf{0}}

\newcommand{\divergence}{\mathop{\rm div}\nolimits}
\newcommand{\curl}{\mathop{\rm curl}\nolimits}

\newcommand{\p}{\partial}

\newcommand{\scl}{\mathcal{L}}

\usepackage[margin=1in]{geometry}

\date{}
\begin{document}
\title{Field Dislocation Mechanics, Conservation of Burgers vector, and the augmented Peierls model of dislocation dynamics}

\author{Amit Acharya\thanks{Department of Civil \& Environmental Engineering, and Center for Nonlinear Analysis, Carnegie Mellon University, Pittsburgh, PA 15213, email: acharyaamit@cmu.edu.}}

\maketitle
\begin{abstract}
\noindent Dissipative models for the quasi-static and dynamic response due to slip in an elastic body containing a single slip plane of vanishing thickness are developed. Discrete dislocations with continuously distributed cores can glide on this plane, and the models are developed as special cases of a fully three-dimensional theory of plasticity induced by dislocation motion. The reduced models are compared and contrasted with the augmented Peierls model of dislocation dynamics. A primary distinguishing feature of the reduced models is the a-priori accounting of space-time conservation of Burgers vector during dislocation evolution. A physical shortcoming of the developed models as well as the Peierls model with regard to a dependence on the choice of a distinguished, coherent reference configuration is discussed, and a testable model without such dependence is also proposed.

\end{abstract}

\section{Introduction}
This paper is concerned with the extent to which the Peierls model \cite{peierls1940size} of a dislocation arising from slip on a single slip-plane can be approached within the Field Dislocation Mechanics (FDM) theory of continuously distributed dislocations \cite{arora2020unification, zhang2015single}. The Peierls model was extended to dynamics by Eshelby \cite{eshelby1949uniformly,eshelby1956supersonic} and Weertman \cite{weertman1967uniformly}, augmented by Rosakis \cite{rosakis2001supersonic} to allow subsonic steady motion and dissipation, and studied in general conditions of non-steady dislocation motion by Pellegrini \cite{pellegrini2010dynamic,pellegrini2020dynamic,pellegrini2011reply} and Markenscoff \cite{markenscoff2011comment}. FDM models for the elastostatic and elastodynamic response of a body containing a single slip plane of vanishing thickness on which discrete dislocations with continuously distributed cores can glide are developed. The models are compared and contrasted with the augmented Peierls model, which has also been extensively used in the phase field setting, see e.g.~\cite{denoual2004dynamic, wang2010phase} and many other subsequent works (which is beyond the scope of the present work to review).

A brief description of the Peierls(-Eshelby) model is as follows.  We write all vector and tensor components w.r.t. a fixed rectangular Cartesian basis of a system with coordinates $(x,y)$, and note that the $T_{xy}$ and $T_{yy}$ components of the stress tensor form the tangential and the normal traction components on a horizontal plane with unit normal in the positive $y$ direction. The displacement vector is denoted as $\bfu = (u_x, u_y)$. The elastostatic or elastodynamic response of a homogeneous, linear elastic, 2d half-space $y > 0$ is sought, subject to the constraint that the tangential traction on the boundary $y = 0$ is equal to some specified function of the tangential displacement there; i.e.~ $T_{xy}(\nabla u(x, y = 0, t)) = f(u_x(x, y = 0, t))$, where $f$ is a prescribed function, possibly making the problem nonlinear. The displacement field on $y = 0$ is required to satisfy boundary conditions at $x=\pm \infty$. Roughly speaking, for $f$ sinusoidal and $u_x(\cdot, y = 0,t)$ a monotone increasing(decreasing) function of $x$ (and of traveling wave type in dynamics), non-trivial solutions are known to exist. The full problem also involves a lower half-space, with the conditions:
\begin{itemize}
\item the traction is continuous on $y = 0$: ($\lim_{y \to 0} T_{xy}(x, y < 0, t) = \lim_{y \to 0} T_{xy}(x, y > 0, t)$ and similarly for the $T_{yy}$ component;
\item the displacements satisfy $\lim_{y \to 0} u_x(x, y < 0, t) = \lim_{y \to 0} - u_x (x , y > 0, t)$ (antisymmetry of tangential displacements) and continuity of vertical displacements: $\lim_{y \to 0} u_y(x, y < 0 , t) = \lim_{y \to 0} u_y (x , y > 0, t)$.
\end{itemize}

P.~Rosakis \cite{rosakis2001supersonic} showed that there is no dissipation for subsonic steady dislocation motion in the Peierls(-Eshelby) model; since `slow' dislocation motion is believed to be accompanied by dissipation, he augmented the Peierls model to include dissipation with a linear drag contribution in the slip rate.

A fundamental feature of FDM is the utilization of a conservation law for the Burgers vector \cite{mura1963continuous}-\cite[App.~B]{acharya2011microcanonical}, phrased in terms of the evolution of the dislocation density tensor field $\bfalpha$:
\begin{equation}\label{eq:disloc_conserv}
  \p_t \bfalpha = - \curl (\bfalpha \times \bfV).  
\end{equation}
Here $\bfV$ is the dislocation velocity field, constitutively prescribed to satisfy non-negative entropy production in the body \cite{acharya2003driving,acharya2004constitutive}. While the setup of the augmented Peierls dynamics to account for subsonic motion accounts for non-negative dissipation \cite{rosakis2001supersonic}, there is no explicit accounting of Burgers vector conservation in that model as a hard constraint. Defining $ \bfalpha := -\curl \bfU^{(p)}$ where $\bfU^{(p)}$ is the plastic distortion, \eqref{eq:disloc_conserv} implies the following evolution equation for the plastic distortion:
\[
\p_t \bfU^{(p)} = -\curl \bfU^{(p)} \times \bfV,
\]
(up to a gradient of a vector field that cannot be unambiguously associated with the motion of dislocations, and therefore assumed to vanish herein).
This theoretical structure, arising from the conservation of Burgers vector (in space-time regions without dislocation flow on their boundary), leads to an important difference in the structure and predictions of the resulting reduced models in comparison to the augmented Peierls model, and the present work is essentially a detailed explanation of this fact.

\section{Derivation of an ansatz within FDM for vanishing interplanar spacing}
\begin{figure}
\centering
\includegraphics[width=5.0in,height=3in]{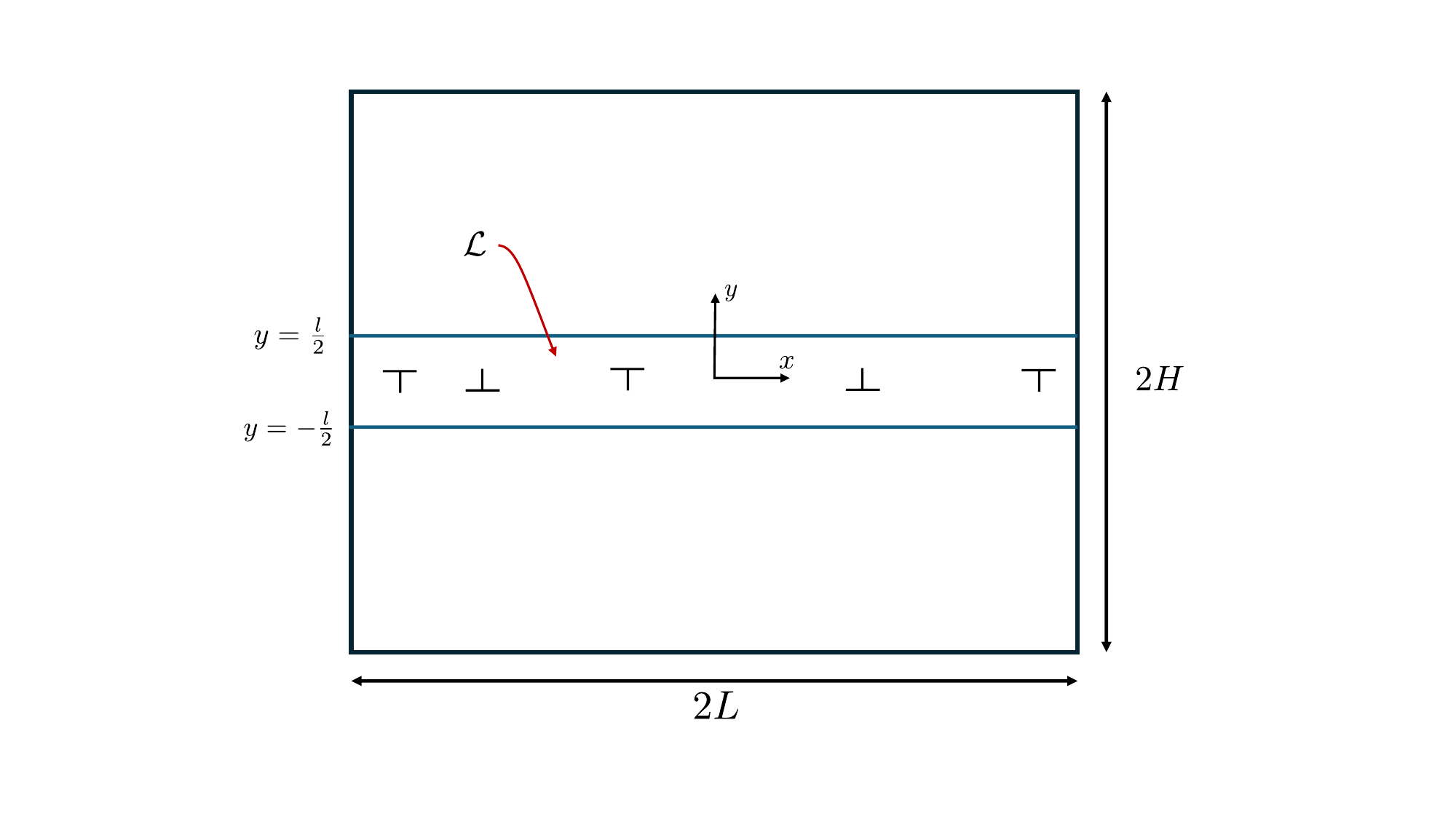}
\caption{Schematic of a body with a layer in which slip occurs between two atomic planes. Edge dislocations move in the layer. Of interest are models of the mechanical response in the limit $l \to 0$.}
\label{fig:fault_layer}
\end{figure}

With reference to Fig.~\ref{fig:fault_layer}, 
and as studied in detail in \cite{zhang2015single,zhang2017continuum} and extended to the study of geophysical rupture in \cite{arora2024emergent}, consider a domain  $\Omega = (-L, L) \times (-H,H)$ in which there is a strip $\scl = (-L, L) \times \left(\frac{l}{2}, \frac{l}{2}\right)$; the width $l$ physically represents the inter-(slip)planar separation in a crystal. We will refer to generic spatial coordinates interchangeably as $(x,y) \equiv (x_1,x_2)$, and time as $t$. Considering a displacement field $\bfu = (u_1, u_2): \Omega \times [0,t_f] \to \R^2$ of a reference configuration $\Omega$ and a plastic distortion field, $\bfU^{(p)}$, of the form
\begin{equation}\label{eq:phi_ans}
    \begin{aligned}
        & \phi: (-L,L) \times [0,t_f] \to \R;  \qquad \bfU^{(p)}: \Omega \times [0,t_f] \to \R^{2 \times 2}\\
        & U^{(p)}_{ij}(x,y,t)  = \begin{cases}
            \phi(x,t) \Big(H\left(y+\frac{l}{2}\right) - H\left(y - \frac{l}{2}\right) \Big) \mbox{ for } i = 1 \mbox{ and } j = 2\\
            0 \qquad \mbox{ for } i \neq 1 \mbox{ or } j \neq 2,
        \end{cases}
        \end{aligned}
\end{equation}
where $H$ is the Heaviside function, the stress tensor $\bfT: \Omega \times [0,t_f] \to \R^{2 \times 2}$ is given by
\begin{equation}\label{eq:stress_const}
\bfT = \lambda \ tr\left(\bfE - \bfE^{(p)}\right) I + 2 \mu \ \left(\bfE - \bfE^{(p)}\right), \qquad \left( T_{ij} = \lambda \left(E_{kk} - E^{(p)}_{kk} \right) \delta_{ij} + 2 \mu \left(E_{ij} - E^{(p)}_{ij} \right) \right),
\end{equation}
where $\lambda, \mu$ are the Lam\'e constants, $\bfE = (\nabla \bfu)_{sym}$ is the total strain and $\bfE^{(p)} = \bfU^{(p)}_{sym}$ is the plastic strain. When convenient for compact representation, we will express the isotropic linear elastic constitutive equation for stress as
\[
\bfT = \mathbb{C} \left(\bfE - \bfE^{(p)}\right), \qquad \qquad \left(T_{ij} = C_{ijkl} \left(E_{kl} - E^{(p)}_{kl} \right) \right).
\]

The mechanical equations that need to be satisfied in $\Omega \times (0,t_f)$ are
\begin{subequations}\label{eq:FDM}
    \begin{align}
        \divergence \bfT & = \rho \ddot{\bfu} & \big( T_{ij,j} = \rho \ddot{u}_i \Big)\label{eq:FDM_stress}\\
        \dot{\bfU}^{(p)} & = -\curl \bfU^{(p)} \times \bfV, & \left( \dot{U}^{(p)}_{ij} = - e_{jrs} e_{rmn} U^{(p)}_{in,m} V_s \right)\label{eq:FDM_strnrt}
    \end{align}
\end{subequations}
where $\bfV:\Omega \times [0,t_f] \to \R^2$ is a constitutively specified dislocation velocity field and the overhead dot represents a time derivative. The mass density, $\rho:\Omega \to \R_+$ is a specified field. Here, $e_{ijk}$ represents the Cartesian components of the alternating tensor.

The dislocation density field, $\alpha$, is defined as
\[
\bfalpha = - \curl \bfU^{(p)}; \qquad \left( \alpha_{ij} = - e_{jrs} U^{(p)}_{is,r}  \qquad \mbox{row-wise curls of } U^{(p)}\right)
\]
for the assumed ansatz on $U^{(p)}$ above,
\begin{equation}\label{eq:alpha_ans}
    \begin{aligned}
        \alpha_{13} &= - \phi_{,1}\left(H \left(y+\frac{l}{2}\right) - H\left(y - \frac{l}{2}\right) \right)\\
        \alpha_{ij} & = 0 \quad \mbox{ for all other } i,j.
    \end{aligned}
\end{equation}
We also employ the ansatz, for $v^{(d)}: (-L,L) \times [0,t_f] \to \R$,
\begin{equation*}
    \begin{aligned}
        V_1(x,y,t) & = v^{(d)}(x,t) \left(H \left(y+\frac{l}{2}\right) - H\left(y - \frac{l}{2}\right) \right)\\
        V_2(x,y,t)  & = 0.
    \end{aligned}
\end{equation*}
It is then shown in \cite{zhang2015single} that the evolution
\begin{equation}\label{eq:phi_evol}
    \phi_t(x,t) = \phi_x(x,t) \frac{\phi_x(x,t)}{\hat{B}} \left( \tau(x,t) - \frac{\p \hat{\eta}}{\p \phi} (\phi(x,t)) + \hat{\eps} \phi_{xx} (x,t)\right)
\end{equation}
 is consistent with the Second law of thermodynamics, where
 \[
 \tau(x,t) = \frac{1}{l}\int^{l/2}_{l/2} T_{12} (x,y,t) dy
 \]
 and $\hat{\eta}$ is a smooth double-well or periodic function of $\phi$ with wells at multiples of $\bar{\phi}$. Also, $\hat{B},\hat{\eps}$ are dislocation drag and core-energy related material constants, and $\bar{\phi}, \hat{B}, \hat{\eps}$ are all  assumed to possibly dependent on $l$ in this work. 

 In order to obtain a form of the limiting evolution of plastic strain in the above model for $l \to 0$, the ansatz is further augmented as follows:
 \begin{itemize}
     \item $\phi$ is of the form
     \begin{equation}\label{eq:phi_p}
         \phi(x,t) := p(x,t) \frac{b(l)}{l}.
     \end{equation}
     The function $p(x,t)$ with non-dimensional values will be assumed to be bounded in the limit $l \to 0$. The field 
     \begin{equation}\label{eq:slip}
         s(x,t;l) := p(x,t)b(l) = \phi(x,t) l
     \end{equation}
     represents the physical \emph{slip} across the strip $\scl$. We will refer to the function $p$ as the non-dimensional slip.
     \item The function $\hat{\eta}(\phi)$ is of the form
     \[
     \hat{\eta}(\phi) := C(l) \eta\left(\frac{\phi}{\bar{\phi}(l)}\right),
     \]
     where $C$ has physical dimensions of \emph{stress} (or energy per unit volume), and the values of the function $\eta$ are non-dimensional.
     Then, defining $\bar{p}(l) : = \bar{\phi}(l) \frac{l}{b(l)}$, $\hat{\eta}(\phi) = C(l) \eta\left( \frac{p}{\bar{p}(l)}\right)$, 
     \[
     \frac{\p \hat{\eta}}{\p \phi} \big(\phi(p; b(l),l) \big) = C(l) \frac{\p \eta}{\p p} \left( \frac{p}{\bar{p}(l)} \right) \frac{1}{\bar{p}(l)} \frac{l}{b(l)} = \frac{C(l)}{\bar{\phi}(l)}\frac{\p \eta}{\p p} \left( \frac{p}{\bar{p}(l)} \right).
     \]
      \item The function $\hat{\eps}(l)$ is of the form $\hat{\eps}(l) = \eps(l) d^2$, where the function $\eps$ has physical dimensions of \emph{stress}. The constant $d$ has physical dimensions of \emph{length} and governs the horizontal extent of a typical dislocation core.
     \item The function $\hat{B}(l)$ is of the form $\hat{B}(l) = \frac{B(l)}{d^2}$, where the function $B$ has physical dimensions of \emph{stress.time}.
     \item $\bar{p}(l)$ is bounded as $l \to 0$.
     \item An applied, stress field $T^a$ is operative in the body, arising from the application of traction boundary conditions, calculated asssuming no dislocations are present in the body, i.e.~\eqref{eq:FDM_stress} is solved with the given traction boundary condition as a linear elastic problem (in quasi-statics, the traction boundary condition must be statically admissible). The total stress field in the body is the superposition $T = T^a + T^d$ of two fields, where the field $T^d$ is calculated with vanishing traction boundary conditions supplementing \eqref{eq:FDM}. The stress field $T^a$ affects the stress field $T^d$ (through the evolution of the plastic distortion), but not the other way around, as in \cite{pellegrini2023shock} - this is different from the case when nonlinear elasticity is used to describe the bulk elasticity, see \cite{zhang2015single}. The total displacement, similarly, is the superposition of the the displacement fields of the two problems, say, $u = u^a + u^d$. Naturally, these assumptions are valid because of the linearity of the stress-elastic strain relationship \eqref{eq:stress_const} and balance of linear momentum \eqref{eq:FDM_stress}.
 \end{itemize}
 We will be interested in two cases as $l \to 0$:
 \begin{itemize}
     \item Case I: $b(l) \to b_0 > 0$, a constant, as $l \to 0$. Here, $\bar{\phi}(l) \to \infty$ as $l \to 0$, the plastic distortion $U^{(p)}$ becomes singular as the inter-planar spacing vanishes, and the slip is generally non-vanishing but finite.

     In this case, we assume $B(l) = O\left(l^{-1} \right)$,  $C(l) = O\left(l^{-1} \right)$, and $\eps(l) = O\left(l \right)$
     as $l \to 0$.

     \emph{This limit of FDM is the one that corresponds to the Peierls model, up to the important difference resulting from the leading $(p_x)^2$ term in \eqref{eq:p_evol_lim_qs} and \eqref{eq:p_evol_lim_dyn}}, as discussed in Sec.~\ref{sec:concl}. 
     \item Case II: $b(l) \to 0$ as $l \to 0$.\footnote{This limit is not relevant for the comparison with the Peierls model. It is considered because this is a seemingly physically reasonable limit of independent interest (Burgers vector, being a lattice vector, vanishes as  lattice spacing vanishes), and it forms the basis of some rigorous mathematical literature on dislocations - see \cite{muller2014geometric,ginster2019strain} and associated discussion in \cite[Sec.~4.2.1]{arora2020unification}.} Here, $\bar{\phi}(l) \to \bar{\phi}_0$, a constant, as $l \to 0$, the plastic distortion $U^{(p)}$ remains bounded as the inter-planar spacing vanishes, and the slip vanishes.

     In this case, we assume $B(l) = O\left( 1 \right)$,  $C(l) = O\left(1 \right)$, and $\eps(l) = O\left(1 \right)$
     as $l \to 0$.
 \end{itemize}
Then the evolution of the non-dimensional slip function is obtained from \eqref{eq:phi_evol} to be
\begin{equation}\label{eq:p_evol}
    p_t(x,t) = \frac{b(l)}{l\, B(l)}d^2 \, (p_x)^2 (x,t) \, \left( \tau(x,t) - \frac{C(l)}{\bar{\phi}(l)} \, \frac{\p \eta}{\p p} (p(x,t); \bar{p}(l))+ \frac{\eps(l) b(l)}{l} d^2 \, p_{xx}(x,t)\right)
\end{equation}
defined for $(x,t) \in (-L,L) \times [0,t_f]$ where
\begin{equation}\label{eq:def_tau}
\begin{aligned}
    \tau(x,t) & = \tau^d(x,t) + \tau^{a}(x,t),\\
\tau^a(x,t) : = \frac{1}{l} \int^{l/2}_{-l/2} T_{12}^a(x,y,t) \,dy; & \qquad \tau^d(x,t) : = \frac{1}{l} \int^{l/2}_{-l/2} T_{12}^d(x,y,t) \,dy.
\end{aligned}
\end{equation}
For direct comparison with the equation governing the displacement discontinuity field of the augmented Peierls model, all occurrences of the field $p$ can be replaced by $p = \frac{s}{b_0}$, where the slip $s = l \phi$ (\ref{eq:phi_p}-\ref{eq:slip}) is the displacement discontinuity\footnote{In the Peierls model the slip is referred to as $u$, but since our 3-d model involves the displacement field with its canonical name, $u$, we refer to the slip with the letter $s$.} across the slip plane in the limit $l \to 0$. This follows from
\begin{equation*}
\begin{aligned}
   \delta(x,t) & := u_1(x,l/2,t) - u_1(x,-l/2,t) \\
   & = \int_{-l/2}^{l/2} \p_y u_1 (x,y,t) dy = \int_{-l/2}^{l/2} U^{(e)}_{12} (x,y,t) dy + \int_{-l/2}^{l/2} \phi(x,t) dy,
\end{aligned}
\end{equation*}
and noting that the elastic distortion $U^e = \nabla u - U^{(p)}$ is integrable as a function of the $y$ coordinate (and even as a function of $(x,y)$). The displacement difference field, $\delta$, across the slip layer $\scl$  was defined as Eqn.~(6) in \cite{zhang2015single} and referred to as the slip therein because of the correspondence shown here in the limit $l \to 0$.

We note that for the limit represented by Case II, while there is plasticity, dissipation, and the presence of incompatibility characterized by a non-vanishing dislocation density field in this case, there is no displacement discontinuity. Indeed, the dislocation density (up to sign) is given by
\[
\phi_x =  p_x \frac{b(l)}{l},
\]
so that for a dislocation `core' with a spatially inhomogeneous distribution of the nondimensional slip $p(x)$ for $x \in [a,b]$, the Burgers vector of the core is given by the integral of the dislocation density over the core region $(a,b) \times (-l/2, l/2)$:
\[
\sim b(l) \,  l \int_a^b \frac{1}{l} p_x \, dx = b(l) \Big(p(b) - p(a) \Big) \to 0 \qquad \mbox{for Case II.}
\]

\subsection{Quasi-static mechanical force balance}\label{sec:qs}

For specified statically admissible traction boundary conditions, $\bfg$, specified on $\p \Omega$ with outward unit normal field $n$, it can be shown \cite{acharya2019structure} that the stress field, $\bfT = \mathbb{C} (\nabla \bfu - \bfU^{(p)})$ of the problem \eqref{eq:FDM_stress} \emph{for $\rho \ddot{\bfu} \approx 0$} and given by
\begin{equation*}
    \begin{aligned}
        \divergence \big( \mathbb{C} (\nabla \bfu - \bfU^{(p)}) \big) & = 0 \qquad \mbox{ on } \Omega \\
        \big( \mathbb{C} (\nabla \bfu - \bfU^{(p)}) \big) \bfn & = \bfg \qquad \mbox{ on } \p \Omega,
    \end{aligned}
\end{equation*}
can be alternatively obtained by solving the problem
\begin{equation}\label{eq:FDM_alpha_stress}
    \begin{aligned}
        \curl \bfU^{(e)} & = \alpha = - \curl \bfU^{(p)} \qquad \mbox{ on } \Omega\\
        \divergence \big( \mathbb{C} \, \bfU^{(e)}\big) & = 0 \qquad \mbox{ on } \Omega \\
        \big( \mathbb{C}\, \bfU^{(e)} \big) \bfn & = \bfg \qquad \mbox{ on } \p \Omega
    \end{aligned}
\end{equation}
for $\bfT = \mathbb{C} \, \bfU^{(e)}$, i.e., in the traction boundary value problem of linear elasticity with dislocations, the stress is uniquely determined from the dislocation density field $\alpha$ and the applied traction $g$. 

We now assume $\frac{L}{l} \to \infty, \frac{H}{l} \to \infty$, i.e.~an infinitely extended body, to facilitate an `easier' analytical treatment. Then, for $g = 0$ and homogeneous, isotropic linear elasticity with shear modulus $\mu$ and Poisson's ratio $\nu$, the system \eqref{eq:FDM_alpha_stress} can be solved by Kr\"oner's stress function method \cite{kroner1981continuum}, and for a straight (in the $x_3/z$ direction) edge dislocation with Burgers vector in the $x$-direction, the shear stress is given by
\[
T_{12}^d(x,y) = \int^\infty_{- \infty} \int^\infty_{- \infty} \frac{\mu}{2 \pi (1 - \nu)} (x - x') \left( \frac{(x - x')^2 - (y - y')^2} {((x - x')^2 + (y - y')^2)^2}\right) \alpha_{13} (x',y') \, dx' dy'
\]
(where the Green's function for the shear stress in the static problem can be read-off from the integrand).

Thus, for present purposes
\begin{equation*}
    \begin{aligned}
        T_{12}^d(x,y,t) & = \frac{\mu}{2 \pi (1 - \nu)} \int^\infty_{- \infty} \int^\infty_{- \infty} (x - x') \left( \frac{(x - x')^2 - (y - y')^2} {((x - x')^2 + (y - y')^2)^2}\right) \\
         & \qquad \qquad \qquad \qquad \qquad \left(- \phi_{x'} (x',t) \right) \left(H\left(y+l/2\right) - H\left(y - l/2\right) \right)\, dx' dy',
    \end{aligned}
\end{equation*}
so that
\begin{equation*}
    \begin{aligned}
        \tau^d(x,t) & := \frac{1}{l} \int^{l/2}_{-l/2} T^d_{12}(x,y,t) \,dy \\
        & = \frac{- \mu}{2 \pi (1 - \nu)}\frac{b(l)}{l} \int^{l/2}_{-l/2} \int^{\infty}_{-\infty}  \, p_{x'}(x',t) \,\frac{1}{l} \int^{l/2}_{-l/2} (x - x')\frac{(x - x')^2 - (y - y')^2} {((x - x')^2 + (y - y')^2)^2} \, dy' dx' dy.
    \end{aligned}
\end{equation*}
We note that for $x \neq x'$ the integrand of the last integral above is a continuous function of $y$ and $y'$ (for $x = x'$ and for all $y \neq y'$, that integrand is $0$). Consequently, we assume the integrals w.r.t $y,y'$ as $l \to 0$ to yield $\frac{l^2}{(x - x')}$ and hence
\begin{equation}\label{eq:tau_d}
    \lim_{l \to 0} \tau^d(x,t) = \left( \lim_{l \to 0} b(l)\right) \frac{- \mu}{2 \pi (1 - \nu)} \int^\infty_{- \infty} \frac{1}{(x - x')} \, p_{x'}(x',t) \, dx'.
\end{equation}
Evidently,
\begin{itemize}
    \item in Case I where $\lim_{l \to 0} b(l) = b_0 > 0$, the dislocation stress, $\tau^d(x,t)$, on the slip plane is generally non-vanishing;
    \item in Case II where $\lim_{l \to 0} b(l) = 0$, the dislocation stress, $\tau^d(x,t)$, on the slip plane vanishes and the effects of dislocation interaction is absent in the problem for this scaling - see related discussion in \cite[Sec.~4.2.1]{arora2020unification}.
\end{itemize}

The fundamental question of understanding dislocation dynamics on a single slip plane in the limit $l \to 0$ within FDM in the given ansatz is to define a sequence of solutions of \eqref{eq:p_evol} parametrized by $l$ and understand the limit of the sequence as $l \to 0$, \emph{and} discover the evolution equation that such a limit satisfies. This is a challenging question beyond the scope of this paper (and perhaps the current state-of-the-art of PDE, Calculus of Variations, and Nonlinear Analysis).

Instead, as a modest alternative based on the preceding considerations, we conjecture the following evolution equation for the nondimensional slip function $\lim_{l \to 0} \frac{s(x,t;l)}{b(l)} = p(x,t)$: with the definitions
\begin{equation}\label{eq:consts}
    \lim_{l \to 0}\, \frac{b(l)d^2 }{l\, B(l)}=: \mathsf{f}; \qquad \lim_{l \to 0} b(l) =: \mathsf{b}; \qquad \lim_{l \to 0} \frac{C(l)}{\bar{\phi}(l)} =: \mathsf{g}; \qquad \lim_{l \to 0} \frac{\eps(l) b(l) d^2}{l} =: \mathsf{c},
\end{equation}
\begin{equation}\label{eq:p_evol_lim_qs}
\begin{aligned}
    p_t(x,t)  = \mathsf{f} \, (p_x(x,t))^2 \, \bigg(   & \frac{- \mu \mathsf{b}}{2 \pi (1 - \nu)} \int^\infty_{- \infty} \frac{1}{(x - x')} \, p_{x'}(x',t) \, dx' \\
    & + \ \tau^a(x,t) \ - \ \mathsf{g} \, \frac{\p \eta}{\p p}(p(x,t)) \ + \ \mathsf{c} \, p_{xx}(x,t) \bigg).
\end{aligned}
\end{equation}
\subsection{Dynamic mechanical force balance}\label{sec:dyn}
The traction boundary conditions generating the field $T^a$ now need not be statically admissible. The only difference from the result \eqref{eq:p_evol_lim_qs} for the quasi-static case arises from the calculation of $\tau^d$ in \eqref{eq:p_evol}. 

In the following, we will often use the notation $r := (x,y,z)$, $r' := (x',y',z')$, $\bar{r} := (x,y)$, $\bar{r}' := (x',y')$, $d^3r' := dx'dy'dz'$, and $d^2r' := dx'dy'$, and spatial integrals will be over all space, unless otherwise specified. A subscript appearing on the fields $\phi$ or $p$ will represent partial derivatives of these fields w.r.t the subscripts.

Mura \cite[Eqn.~(45)]{mura1963continuous} derived the expression for the elastic distortions:
\begin{subequations}\label{eq:mura_stress}
    \allowdisplaybreaks
    \begin{align}
        U^{(e)}_{mn}(r,t) & =  \int^t_{-\infty} \int - e_{njh} C_{ijkl} G_{km,l}(r-r',t-t') e_{pqh} U^{(p)}_{iq,p}(r') - \rho \dot{G}_{im}(r-r',t-t') \dot{U}^{(p)}_{in}(r',t') \, d^3r'dt' \notag \\
        & \mbox{ (and by renaming indices) }  \notag\\
      U^{(e)}_{ij}(r,t) & = \int^t_{-\infty} \int e_{jln} C_{mkpl} G_{im,k}(r-r',t-t') \alpha_{pn}(r') - \rho \dot{G}_{im}(r-r',t-t') \dot{U}^{(p)}_{ij}(r',t') \, d^3r'dt', \tag{\ref{eq:mura_stress}}  
    \end{align}
\end{subequations}
where $G_{ij} = G_{ji}$ is the Green's function for linear, homogeneous, isotropic elastodynamics deduced by Love \cite{love2013treatise} (presented in modern tensor notation as equation (8) in \cite{mura1963continuous}).

 For our ansatz where $U^{(p)}_{12}$ and $\alpha_{13}$ are the only non-zero components of the plastic distortion and the dislocation density, we have that
    \allowdisplaybreaks
    \begin{align}
        U^{(e)}_{12}(r,t) & = \int^t_{-\infty} \int e_{213} G_{1m,k}(r-r',t-t')C_{mk11} \alpha_{13}(r',t') - \rho \dot{G}_{11}(r-r',t-t')\dot{U}^{(p)}_{12}(r',t') \, d^3r'dt' \notag\\
        U^{(e)}_{21}(r,t) & = \int^t_{-\infty} \int e_{123} G_{2m,k}(r-r',t-t')C_{mk12} \alpha_{13}(r',t')  \, d^3r'dt'. \notag
     \end{align}
We now define
\[
\bar{G}(\bar{r} - \bar{r}',t-t') := \int^\infty_{-\infty} G_{ij}(\bar{r} - \bar{r}', 0 - z', t - t') \, dz'.
\]
Noting that 1) $G_{im,z} = - G_{im,z'}$ and integration by parts of such terms w.r.t $z'$ and evaluation at $z = 0$ eliminates all integrations w.r.t $z'$ and derivatives w.r.t~$z,z'$ from the expressions of $U^{(e)}_{ij}(x,y,0,t), ($ $i,j = 1,2$), and 2) the primary dependence of $U^{(p)}_{12}$ and $\alpha_{13}$ on only the $x$-coordinate, and their possibly non-vanishing nature only in $\scl$, we have that
\[U^{(e)}_{12}(\bar{r},0,t) = \int_{-\infty}^t\int_{-\infty}^{\infty}\int_{-l/2}^{l/2} \bar{G}_{1m,k}(\bar{r} - \bar{r}',t-t')C_{mk11} \phi_{x'}(x',t') - \rho \dot{\bar{G}}_{11}(\bar{r} - \bar{r}', t - t') \phi_{t'}(x',t') \, dy' dx' dt',
\]
for $k$ running over $1,2$. Finally, noting that
\[
C_{ijkl} = \lambda \delta_{ij}\delta_{kl} + \mu (\delta_{ik}\delta_{jl} + \delta_{il} \delta_{jk})
\]
and using \eqref{eq:phi_p} we obtain
    \allowdisplaybreaks
    \begin{align}
        U^{(e)}_{12}(\bar{r},0,t) & = \frac{b(l)}{l}\int_{-l/2}^{l/2} \int^t_{-\infty} \int_{-\infty}^{\infty} \Big( (\lambda + 2 \mu) \bar{G}_{11,1} + \lambda \bar{G}_{12,2}\Big)(\bar{r} - \bar{r}', t - t') \, p_{x'} (x',t') \notag\\
        & \qquad \qquad \qquad \qquad \qquad - \rho \dot{\bar{G}}_{11} (\bar{r} - \bar{r}', t- t')\, p_{t'}(x',t') \, dx'dt'dy' \notag\\
        U^{(e)}_{21}(\bar{r},0,t) & =  - \frac{b(l)}{l}\int_{-l/2}^{l/2} \int^t_{-\infty} \int_{-\infty}^{\infty} \mu \Big( \bar{G}_{21,2} + \bar{G}_{22,1} \Big) (\bar{r} - \bar{r}', t - t') \, p_{x'} (x',t')  \, dx'dt'dy'. \notag
     \end{align}
Now,
\[
T^d_{12}(x,y,0,t) = \mu \big( U^{(e)}_{12} + U^{(e)}_{21} \big) (x,y,0,t),
\]
and assuming a regularization of the Green's function as in Pellegrini \cite{pellegrini2020dynamic}, so that it can be assumed to be a continuous function, $\tau^d(x,t)$ (see, \eqref{eq:def_tau}) in this elastodynamic case is given by
\begin{subequations}\label{eq:tau_dd}
    \allowdisplaybreaks
    \begin{align}
\tau^{dd}(x,t) & = \mu^2 b(l) \int^t_{-\infty}\int_{-\infty}^{\infty} \Big( \omega^2 \bar{G}_{11,1}  -  \bar{G}_{22,1} +  (\omega^2 - 3) \bar{G}_{12,2} \Big)(x - x', 0,t - t') \, p_{x'} (x',t') \, dx' dt'\notag\\
& \qquad \qquad \qquad \qquad \quad - \frac{1}{c_s^2}  \dot{\bar{G}}_{11} (x - x', 0,t - t') \, p_{t'} (x',t') \, dx' dt' \tag{\ref{eq:tau_dd}
}
  \end{align}
\end{subequations}
where $\omega := \frac{c_l}{c_s}$, with $c_l := \sqrt{\frac{\lambda + 2 \mu}{\rho}}$ the dilatational wave speed, and $c_s := \sqrt{\frac{\mu}{\rho}}$ the shear wave speed.

The corresponding equation for the elastodynamic nondimensional slip function in the limit $l \to 0$ may be expected to be given by (with constants defined in \eqref{eq:consts})
\begin{equation}\label{eq:p_evol_lim_dyn}
\begin{aligned}
   &  p_t(x,t) = \mathsf{f} \, (p_x)^2(x,t) \, \bigg(   \tau^{D}(x,t) \ + \ \tau^a(x,t) \ - \ \mathsf{g} \, \frac{\p \eta}{\p p}(p(x,t)) \ + \ \mathsf{c} \, p_{xx}(x,t) \bigg); \qquad \tau^{D}(x,t) := {\mathsf b} \frac{\tau^{dd}(x,t)}{b(l)}\\
   & \Longleftrightarrow\\
   & \int_{-\infty}^t\int_{-\infty}^{\infty} \left( \delta(x-x',t-t') \ + \  \frac{{\mathsf{b \, f}} \, \mu^2}{c_s^2} \, (p_x)^2(x,t) \  \dot{\bar{G}}_{11} (x - x', 0,t - t') \right)\, p_{t'} (x',t') \, dx' dt'\\
   & =  \mathsf{f} \, (p_x)^2(x,t) \, \bigg( \int^t_{-\infty}\int_{-\infty}^{\infty} \mu^2 {\mathsf b} \Big( \omega^2 \bar{G}_{11,1} -  \bar{G}_{22,1} +  (\omega^2 - 3) \bar{G}_{12,2}  \Big)(x - x', 0,t - t') \, p_{x'} (x',t') \, dx' dt' \\
   & \qquad \qquad \qquad \qquad \qquad \qquad \qquad + \ \tau^a(x,t) \ - \ \mathsf{g} \, \frac{\p \eta}{\p p} \big(p(x,t)\big)\ + \ \mathsf{c} \, p_{xx}(x,t) \bigg).
\end{aligned}
\end{equation}
Much like the quasistatic case, in the dynamic case the dislocation interaction through the stress field $\tau^D$ vanishes for Case II. 

As in the case of quasistatic force balance Sec.~\ref{sec:qs}, for the limit Case I, the material constants $\mathsf{f}, \mathsf{b}, \mathsf{g}, \mathsf{c}$ affect the $p$ evolution, and through it the dislocation stress field $\tau^d$.

\section{Discussion and concluding remarks}\label{sec:concl}
Within a formal ansatz of small elastic, plastic, and total distortions about a given reference configuration, a model of single slip induced by the motion of edge dislocations is developed within Field Dislocation Mechanics theory, in the limit of vanishing inter-planar lattice spacing. Both quasi-static and dynamic mechanical force balances are considered, with the aim of understanding the similarities and differences between the resulting model and the augmented Peierls model in similar settings \cite{peierls1940size,eshelby1949uniformly, weertman1967uniformly, rosakis2001supersonic, pellegrini2020dynamic}. The following conclusions may be drawn:
\begin{enumerate}
    \item The dynamics of dislocations in FDM is strongly constrained by the statement of conservation (in space-time) of the Burgers vector, which forms one of its fundamental field equations. The Peierls model (and its various descendants) does not involve such a `hard' constraint. This conceptual difference directly results in the main difference between the models descended from FDM and the Peierls model: in \eqref{eq:p_evol_lim_qs} and \eqref{eq:p_evol_lim_dyn}  if the material constant $d$ in $\mathsf f$ were to be replaced by $(p_x)^{-1}$ (definitely not a constant!) for $p_x \neq 0$, with ${\mathsf f} (p_x)^2 = 1$ when $p_x = 0$, one obtains the augmented Peierls model\footnote{In the full (as opposed to the scalar)  system of FDM, whether in one or multiple space dimensions, the `multiplicative' presence of the dislocation density $\alpha = - \curl U^{(p)}$ cannot be obliterated by the device employed here.} proposed first by Rosakis \cite{rosakis2001supersonic} and generalized to non-steady dislocation motions and extensively analyzed and elucidated by Pellegrini \cite[and references therein]{pellegrini2010dynamic,pellegrini2020dynamic}, including making connections with the work of Mura \cite{mura1963continuous}. However, as they stand, the FDM limiting equations \eqref{eq:p_evol_lim_qs} and \eqref{eq:p_evol_lim_dyn} are degenerate parabolic transport equation with strong wave-propagative features away from the cores of dislocations, whereas the augmented Peierls problem is a nonlinear reaction-diffusion type equation (and in the system, as opposed to the scalar case, even in 1 space dimension, the dissipative dynamics is very different from that corresponding to a system of reaction diffusion equations, and even from a system of hyperbolic conservation laws in a subtle respect, as explained in \cite[Sec.~5]{acharya2011equation}). Further physical differences are discussed in remarks 2 and 3 below.
    \item The fundamental physical implication of involving the conservation of Burgers vector as a field equation is the statement that plastic strain rate at a field point in space-time in FDM can arise \emph{only if} there exists a dislocation (as characterized by a non-zero $p_x$) at that field point, regardless of the magnitude of strain energy or stress available as driving force, whether in quasi-static or dynamic settings. 
    
    As a result, in FDM, dissipation can only occur in the core - this is not so in the augmented Peierls model (whether for steady-state or non-steady dislocation motion) or its phase field generalizations.

    Furthermore, in dynamics with inertia, because a spatially homogeneous, but time-dependent plastic strain history affects the elastic distortion field in the entire body \eqref{eq:mura_stress}-\eqref{eq:tau_dd} (and not only in the slip plane), the elastic distortion, and consequently the stress fields, of the two models will in general also be different. This is a well-understood fact about the difference between the elasto\emph{dynamic} stress and elastic distortion fields of a plastic distortion field and its corresponding dislocation density field (if any) \cite{acharya2019structure, willis1965dislocations} under traction boundary conditions, a difference that is not present in the corresponding elastostatic fields. 
    
    To elaborate in the present context, consider a `source' point, say $r'$ which undergoes a locally spatially homogeneous, nontrivial plastic strain history, i.e., $p_t(r',\cdot) \neq 0$ for some interval of time with $p_x(r',\cdot) = 0$ for all time. Then the elastic distortion field at a field point $r$ of the body in the two models will be different due to the presence of such source points in the body. Moreover, such plastic slip histories are quite possible - one example is a transient, spatially homogeneous slip in a layer which cannot not arise from the motion of dislocations\footnote{The conservation of Burgers vector does allow, in principle, an additive plastic strain rate field in the strict form a gradient of a vector field which can represent such layer-homogeneous transient slip fields, if desired. However, such plastic strain rate fields at any fixed instant of time cannot arise from the motion of localized dislocation cores.}.
    \item As studied in detail in \cite{zhang2015single}, the `multiplicative' $p_x^2$ in the FDM evolution for plastic strain \eqref{eq:p_evol_lim_qs} most likely results in the occurrence of a Peierls stress even in a PDE model which is translationally-invariant (a feature that is absent in the original Peierls model). This a falsifiable conjecture, ripe for rigorous mathematical study.
    \item The Peierls model and its dissipative descendants are not derived from a general theory of a three dimensional body and what that generalization should be is by no means obvious, particularly as a general theory of continuum mechanics with no kinematic restrictions. This is not the case with FDM and, indeed, the three-dimensional generalization of the evolution for the plastic distortion \eqref{eq:FDM_strnrt} involves fundamentally different characteristics than its slip-plane counterpart \eqref{eq:p_evol}.
    \item Finally, as useful as the models \eqref{eq:p_evol_lim_qs} and \eqref{eq:p_evol_lim_dyn} can be as shown by the related studies in \cite{zhang2015single, zhang2017continuum, } of the model \eqref{eq:phi_evol}, there is a fundamental physical shortcoming of such models (as well as the augmented Peierls model and descendants, also see \cite{baggio2023inelastic}) arising from the crucial dependence of their energy densities on the notion of slip or plastic strain, which cannot be physically defined without resort to a reference configuration for the total and plastic deformations. However, given a crystalline body with dislocations, there is no natural way in which a distinguished global, coherent reference configuration can be associated with it, in practice or in physically faithful modeling, say in a given atomistic assembly.

    I view this as a rather serious physical shortcoming of such models, even though a difficult one to overcome in PDE models since the whole device of inducing a localized core of a dislocation in all such models up to this point depends crucially on introducing a non-convex energy contribution in the slip with vanishing or low-energy preferred states. The total displacement gradient incurs these states as `stress-free' states, up to kinematic constraints arising from Dirichlet boundary conditions or compatibility, the latter providing the correct constraints for predicting dislocation stress fields.

    With that as motivation, in the following we provide some speculation on a falsifiable model w.r.t its capabilities in predicting dislocation behavior in elastic solids which does not have the conceptual shortcoming described above; another option is described in \cite{acharya2020field}. For a domain $\Omega \subset \R^3$ and $[0,t_f] \in \R$ an interval of time, with the basic fields
\allowdisplaybreaks
    \begin{align}
        \bfv&: \Omega \times [0,t_f] \to \R^3 \qquad &\mbox{ material velocity} \notag\\
        \bfT &: \Omega \times [0,t_f] \to \R^{3 \times 3}_{sym} \qquad &\mbox{ stress} \notag\\
       \bfU \equiv  \bfU^{(e)}&: \Omega \times [0,t_f] \to \R^{3 \times 3} \qquad &\mbox{ elastic distortion} \notag\\
        \bfalpha \equiv \curl \bfU^{(e)} &: \Omega \times [0,t_f] \to \R^{3 \times 3} \qquad &\mbox{ dislocation density} \notag\\
        \bfV&: \Omega \times [0,t_f] \to \R^3 \qquad &\mbox{ dislocation velocity} \notag\\
        \rho&: \Omega \times [0,t_f] \to \R_+ \qquad &\mbox{ mass density (specified)}, \notag
    \end{align}
\begin{equation*}
\begin{aligned}
     \mathbb{C} &: \R^{3 \times 3}_{sym}  \to \R^{3 \times 3}_{sym} \qquad &\mbox{elastic moduli} \\
     \mathbb{M}\big|_\alpha &: \R^3 \to \R^3 \qquad & \mbox{positive semi-definite dislocation mobility tensor}\\
     & & \mbox{ possibly dependent on } \alpha,
\end{aligned}
\end{equation*}
the governing equations are given by
\begin{subequations}\label{eq:gov_eq}
    \begin{align}
        \divergence \bfT &= \begin{cases}
            \rho \, \p_t \bfv \qquad \mbox{ or}\\
            0 \qquad \mbox{ quasi-static force balance} 
        \end{cases} \label{eq:blm}\\
       \nabla \bfv - (\curl \bfU) \times \bfV   & = \p_t \bfU \label{eq:compat}.
    \end{align}
\end{subequations}
The stress and dislocation velocity are constitutively specified as
\[
\bfT = \mathbb{C} \bfU,
\]
and, with
\[
\alpha \to \eta(\alpha): \R^{3 \times 3} \to \R
\]
a generally nonconvex function of the dislocation density,
\begin{equation}\label{eq:disloc_vel}
    \bfV = \mathbb{M}\big|_\alpha \bfX:\left( \left( \bfT + \curl \frac{\p \eta}{\p \bfalpha} - \curl (\kappa \Delta \bfalpha)  \right)^T \bfalpha \right).
\end{equation}
Here $(X:(AB))_i = e_{ijk}A_{jr}B_{rk}$ for any $3 \times 3$ matrices $A,B$, and $\kappa \in \R_+$ is a `small' material constant.

The energy density $\eta$ is intended to induce energetic constraints favoring dislocations of specific characters particular to a specific crystallography and the magnitude of their Burgers vector. A simplest illustrative example of $\eta$ is of the form shown in Fig.~\ref{fig:eta}.

\begin{figure}[htb]
\centering
\includegraphics[width=3in,height=3in]{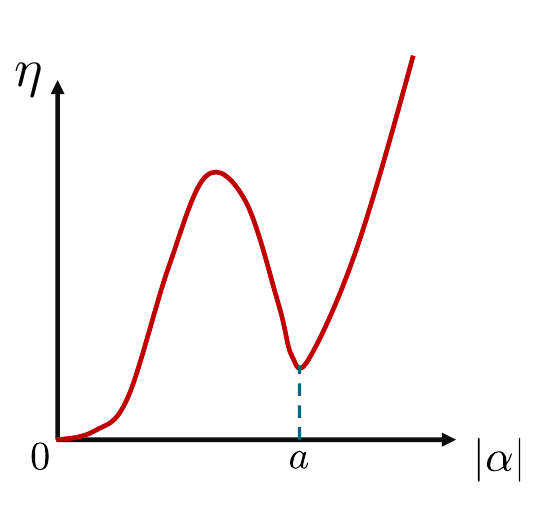}
\caption{Schematic of a simplest energy density contribution due to dislocations. The energy prefers either no dislocations or dislocation density fields with pointwise magnitude $a = \frac{b}{\xi^{\parallel}\xi^\perp}$ where $b$ is the magnitude of the Burgers vector, $\xi^{\parallel}$  is the typical dimension of the core parallel to the slip plane, and $\xi^\perp$ the typical core height orthogonal to the slip plane.}
\label{fig:eta}
\end{figure}

The following facts about such a model are worth noting:
\begin{itemize}
    \item We note that \eqref{eq:compat} implies the conservation of Burgers vector (in space-time) given by
\[
\p_t \bfalpha = - \curl (\bfalpha \times \bfV).
\]
\item The function $\eta$ and the elastic moduli $\C$ are unambiguously physically specifiable. The term involving $\kappa$ is physically dubious, but a mathematically necessary regularization for the model; here we mention, however, the nascent possibility of avoiding such mathematically necessary higher-order regularizations by using solution concepts and techniques \cite{ach_HCC} that can possibly extract physically meaningful behavior from what are conventionally considered `ill-posed' models, a case study shown in \cite{singh2024hidden}.
\item The theory guarantees that, defining the stored energy density
\begin{equation}\label{eq:new_energy}
    \psi = \half \bfU:\mathbb{C} \bfU + \eta(\bfalpha) + \half \kappa |\nabla \bfalpha|^2,
\end{equation}
non-negative dissipation
\[
\int_{\p \Omega} \Big( (\C \bfU) \bfn \Big)\cdot \bfv \, da - \frac{d}{dt} \left( \int_\Omega \,\half \,\rho |\bfv|^2 + \psi \  d^3x \right) \geq 0,
\]
i.e., the power of applied forces minus the rate of change of the kinetic and stored energy of the body is non-negative, holds for all trajectories $(v,U)$ satisfying \eqref{eq:gov_eq}. Here, $n$ is the outward unit normal field on $\p \Omega$. While we have chosen the elastic energy contribution to be quadratic for simplicity, it could very well be nonlinear in the elastic strain.
\item While the proposed model is at small deformations, such a model of defect mechanics lends itself to immediate generalization to finite deformations as in \cite{arora2020unification} allowing also for homogeneous dislocation nucleation \cite{garg2015study}, as well as the inclusion of couple-stresses and  interfacial defects \cite{AcharyaFressengeas2015}.
\item An intriguing question is whether a stored energy density of the form $\hat{\psi} = \Phi(\bfU_{sym}) + \bar{\eta}(\bfalpha)$, with $\Phi$ a multiple well energy in the (generally incompatible) elastic distortion and $\bar{\eta}$ a convex regularization, along with (\ref{eq:gov_eq}-\ref{eq:disloc_vel}) with $\bfT = \p_{\bfU_{sym}} \Phi$,  can also be a useful PDE model as compared to the more complicated dependence on the dislocation density discussed in \eqref{eq:new_energy} in the stored energy along with a simpler dependence on the elastic distortion. The former ($\hat{\psi}$ model) has the disadvantage that in the limit of no dislocations, it leads to an ill-posed dynamical model, at least with classical notions of ill-posedness. It remains to be seen what regularization to dynamical behavior, if any, is provided by explicitly allowing for dislocations in the model as described. It also has the somewhat physically unpleasant feature that given a metastable equilibrium configuration of atoms comprising a body with dislocations (with the atoms indistinguishable from one another), away from the cores the elastic distortion of a local neighborhood of atoms would, in the most natural sense, seem to be in the neighborhood of the $\bfU = \bfzero$ state; however, to predict the fields of a dislocated state, a (segment of a) layer of atoms would certainly need to be in an energy well to which $\bfU_{sym} = \bfzero$ does not belong (a state which can, nevertheless, be \emph{theoretically} conceived in a dislocated lattice).
\end{itemize}
    
\end{enumerate}

\printbibliography
\end{document}